\documentclass[pre,floats,twocolumn,showpacs,floatfix,superscriptaddress]{revtex4}
\usepackage{graphics,graphicx,dcolumn,bm,epic,eepic,float}
\usepackage{amssymb,amsmath,multirow,rotate,color,float,mciteplus}
\usepackage[latin1]{inputenc}
\usepackage{psfrag}
\usepackage[dvips]{epsfig}

\usepackage{epstopdf}

\begin{document}

\title{`Gas cushion' model and hydrodynamic boundary conditions for superhydrophobic textures.}

\author{Tatiana V. Nizkaya}
\affiliation{A.N.~Frumkin Institute of Physical
Chemistry and Electrochemistry, Russian Academy of Sciences, 31
Leninsky Prospect, 119071 Moscow, Russia}
\author{Evgeny S. Asmolov}
\affiliation{A.N.~Frumkin Institute of Physical
Chemistry and Electrochemistry, Russian Academy of Sciences, 31
Leninsky Prospect, 119071 Moscow, Russia}
\affiliation
{Central Aero-Hydrodynamic Institute, 140180
Zhukovsky, Moscow region,  Russia}
\affiliation
{Institute of Mechanics, M. V. Lomonosov Moscow State University, 119991 Moscow,
Russia}
\author{Olga I. Vinogradova}
\email[Corresponding author: ]{oivinograd@yahoo.com}
\affiliation{A.N.~Frumkin Institute of Physical
Chemistry and Electrochemistry, Russian Academy of Sciences, 31
Leninsky Prospect, 119071 Moscow, Russia}
\affiliation{Department of Physics, M. V. Lomonosov Moscow State University, 119991 Moscow, Russia}
\affiliation{DWI - Leibniz Institute for Interactive Materials, RWTH Aachen, Forckenbeckstra\ss e 50, 52056 Aachen,
  Germany}

\begin{abstract}

Superhydrophobic Cassie textures with trapped gas bubbles reduce drag, by generating large effective slip, which is important for a variety of applications that involve a manipulation of liquids at the small scale.  Here we discuss how the dissipation in the gas  phase of textures modifies their friction properties. We propose an operator method, which allows us the mapping of the flow in the gas subphase to a local slip boundary condition at the liquid/gas interface. The determined uniquely  local slip length depends on the viscosity
contrast and underlying topography, and can be immediately used to evaluate an effective slip of the texture. Besides superlubricating Cassie surfaces our approach is valid for rough surfaces impregnated by a low-viscosity `lubricant', and even for Wenzel textures, where a liquid follows the surface relief. These
results provide a framework for the rational design of textured surfaces for numerous applications.

\end{abstract}

\pacs {83.50.Rp, 47.61.-k, 68.03.-g}

\maketitle

\section{Introduction.}
Superhydrophobic (SH) textures have raised a considerable interest and motivated numerous studies during the past decade. Such surfaces in the Cassie
state, i.e., where the texture is filled with gas, can induce  exceptional wetting properties~\cite{quere.d:2008} and, due to their superlubricating potential~\cite{bocquet2007,rothstein.jp:2010,vinogradova.oi:2012,voronov.rs:2008}, are also extremely important in context of fluid dynamics. To quantify the drag reduction
 associated with two-component (e.g., gas and solid) SH surfaces with given area fractions it is convenient to construct the effective slip boundary condition (on the scale larger than the pattern characteristic length) for the averaged velocity field. This condition is applied at the imaginary smooth homogeneous surface~\cite{vinogradova.oi:2011,Kamrin_etal:2010}, which mimics the actual one and fully characterizes the flow at the real surface and is generally a tensor~\cite{stone2004,Bazant08}. Once eigenvalues of the slip-length tensor, which depend on both the hydrodynamic boundary condition at the solid/liquid interface and viscous dissipation in the gas phase, are determined, they can be used to solve complex hydrodynamic problems without tedious calculations. A key difficulty is that there is no general analytical theory that relates this dissipation to the relief of the texture, so that  prior work often neglected it, by imposing idealized shear-free boundary conditions at the gas sectors~\cite{philip.jr:1972,priezjev.nv:2005,lauga.e:2003}.

  \begin{figure}[h]
\includegraphics[width=0.22\textwidth]{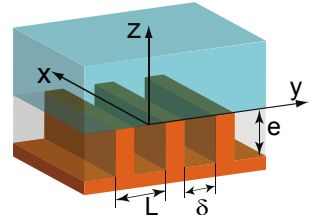}
\caption{Sketch of the typical 1D SH surface represented by rectangular grooves rigorously studied here, but some of our conclusions are general and apply for any 1D and 2D textured surfaces - pillars,
holes, lamelae.}
\label{sketch}
\end{figure}
To account for a dissipation within the gas subphase it is necessary to solve Stokes equations by applying conditions
\begin{equation}
 z=0: \; \mathbf{u}= \mathbf{u}_g, \; \mu \dfrac{\partial  \mathbf{u}_\tau}{\partial z}=\mu_g\dfrac{\partial \mathbf{u}_{g\tau}}{\partial z},
\label{BCcon}
\end{equation}
where $\mathbf{u}$ and $\mu $ are the velocity and the dynamic viscosity of the liquid, and $\mathbf{u}_g$ and $\mu _g$ are those of the gas, $\mathbf{u}_\tau=(u_x,u_y)$  is the tangential velocity.  Although this  problem has been resolved numerically for rectangular grooves~\cite{maynes2007,ng:2010}, such a strategy appears rather hopeless in context of exact analytical results, especially for complex configurations, which are typical for many applications. An elegant semi-analytical approach based on an assumption of a  constant shear inside the groove has been proposed recently~\cite{shoenecker.c:2014}. Note that although this derivation made a significant step forward, it remains approximate and does not take into account the total dissipation in the gas subphase.

To bypass this problem, it is advantageous to  replace the two-phase approach, by a single-phase problem  with spatially dependent partial slip boundary condition~\cite{belyaev.av:2010a,bocquet2007}, which takes a form
\begin{equation}
 z=0: \; \mathbf{u}_\tau-b(x, y)\dfrac{\partial \mathbf{u}_\tau}{\partial z}=0,
\label{BCslip}
\end{equation}
where $b(x, y)$ is the \emph{local} slip length at the gas areas, which is normally assumed to conform the texture relief according to predictions of the `gas cushion' model~\cite{vinogradova.oi:1995a}
\begin{equation}\label{bgas}
b^{x,y} (x, y)\simeq k^{x,y}\dfrac{\mu }{\mu_g} e(x, y),
\end{equation}
where prefactors $k^{x,y}=1$ can reduce to $1/4$ if the net gas flux becomes zero (due to end walls)~\cite{nizkaya.tv:2013}.
Such an approach, justified for a continuous gas layer at a homogeneous surface~\cite{vinogradova.oi:1995a} and later for shallow grooves~\cite{nizkaya.tv:2013}, is by no means obvious for an arbitrary texture, where the gas subphase can be deep and strongly confined.  In such a situation it remains largely unknown if the gas flow can be indeed excluded from the analysis being equivalently replaced by  $b(x, y)$, and how (and whether) this local slip profile is uniquely related to the relief of the texture.

In this article, we propose a general theoretical method, which allows us 
to generalize  the `gas cushion' model for any 1D and 2D two-phase SH textures in the Cassie state, or rough surfaces impregnated by a low-viscosity `lubricant'. We also show that our approach can be applied even for textures in the one-phase Wenzel state, where the liquid follows the topological variations of the texture.

\section{Theory.}

\subsection{General consideration}

To illustrate our approach,  we
consider 1D SH surface of period $L$, and assume the interface to be flat with no
meniscus curvature (see Fig.~\ref{sketch}). Such an idealized  situation, which neglects an additional mechanism for a dissipation due to a meniscus~\cite{sbragaglia.m:2007,harting.j:2008}, has been considered in most
previous publications~\cite{priezjev.nv:2005,ybert.c:2007,vinogradova.oi:2011} and observed in recent experiments~\cite{karatay.e:2013}.  We then impose no-slip at the solid area, i.e. neglect slippage of liquid~\cite{vinogradova.oi:2003,charlaix.e:2005,vinogradova.oi:2009,joly.l:2006} and gas~\cite{seo.d:2013} past hydrophobic surface, which is justified provided the nanometric slip is
small compared to parameters of the texture.
No further assumptions are made, aside from distinguishing between longitudinal and transverse gas flow, with $k^{x}=1$ and $k^{y}=1/4$, to address the most anisotropic case.

The linearity of Stokes
equations implies that the boundary condition at the liquid/gas interface for longitudinal and transverse directions can be
formulated as:
\begin{equation}
 z=0: \; \dfrac{\partial \mathbf{u}_{g\tau}}{\partial z}-\mathbf{P}^{x,y}[\mathbf{u}_{g\tau}]=0,
\label{PS1}
\end{equation}
where we introduced the linear
operator $\mathbf P^{x,y}$ that belongs to a general class of Dirichlet-to-Neumann (DtN) ones~\cite{quarteroni.a:1999}.
The meaning of  Eq.(\ref{PS1}) becomes clear if we project both the velocity and its normal derivative on a grid $\{y_i\}_{i=1}^{N}$ at the liquid/gas interface. Then the operator becomes a matrix $P_{ij}$ that relates the shear rate at a given point $i$ with velocities in every other points of the interface $j=1..N$ (the condition is essentially \emph{nonlocal}).
Unlike the local slip length, the operator depends only on the texture relief, but not on the solution outside. It is universal and, once calculated for a given topography, can be applied for any geometry of the outside flow and any viscosity ratio.
Then, in view of Eq.(\ref{BCcon}), the non-local boundary condition for fluid flow past a SH surface reads:
\begin{equation}
\dfrac{\partial u_{x,y}(y_i,0)}{\partial z}-\dfrac{\mu}{\mu_g}\sum_{j=1}^{N} P_{ij}^{x,y}u_{x,y}(y_j,0)=0,\; i=1..N.
\label{PS_grid}\end{equation}
This boundary condition allows to solve the Stokes equations for the liquid phase separately and to determine the local slip length by using Eq.(\ref{BCslip}):
\begin{equation}
b^{x,y}(y)=\dfrac{\mu}{\mu_g}\dfrac{u_{x,y}(y,0)}{\mathbf{P}^{x,y}[u_{x,y}]}.
\label{localB}
\end{equation}
We recall that he local slip length may depend not only on the texture relief, but also on the state of the liquid phase, which could affect the velocity $u_{x,y}$. However, for a single surface we consider here the local slip length is uniquely related to the texture relief and the generalization of the `gas cushion' model can be constructed. For confined  configurations (e.g. flow in a thin channel) the local slip length will of course be a property of the whole system, but note that the $\mathbf P^{x,y}$-operators will remain exactly the same.

To calculate the matrices $P_{ij}^{x,y}$ we should solve the problem in the gas phase and extract the normal derivative of the solution either analytically or numerically. In this paper we rigorously calculate them for a rectangular groove using a Fourier method. For an arbitrary 1D geometry $\mathbf{P}^{x,y}$ can be expressed in the form of a boundary integral operator involving Green's functions for the Stokes flow \cite{Pozrikidis:book}. As a side note, we remark that it can be similarly constructed for 2D surfaces, but of course by using Green's functions for 3D Stokes flow~\cite{Pozrikidis:book}.
Note however that one does not
expect the main physical picture to be altered in these (more technically challenging) situations, and we leave the study of these complex geometries for a future work.

\subsection{Periodic rectangular grooves.}

For an initial application of our approach,  we consider now periodic rectangular grooves of width $\delta$ and depth $e$. The fraction of gas area is then $\phi =\delta/L$. In this particular case the problem inside the groove can be solved using the Fourier method. For the longitudinal flow this yields an analytical expression for the DtN matrix:
\begin{equation}
\begin{array}{ll}
P_{ij}^{x}=\dfrac{1}{\delta}\left(F_{im} \Pi_{ml} {F}^{-1}_{lj}\right),
\label{PSrect}\\
F_{im}=\cos(k^*_m y_i/\delta), \; \Pi_{ml}=k^*_m\coth \left(k^*_me/\delta
\right)\delta_{ml},
\end{array}
\end{equation}
where $k_m^*=(2m-1)\pi$ and  $\delta_{ml}$ is the Kronecker delta.
 Note that the matrix
combination inside the brackets in Eq.(\ref{PSrect}) depends only
on the aspect ratio, $e/\delta$ (and the spatial grid used). The same is true for the transverse direction, although the matrix $P_{ij}^{y}$ can be obtained only semi-analytically (see Appendix ~\ref{ap:B}).

Having calculated $P_{ij}^{x,y}$,  we can then use the Fourier method to solve the Stokes equations for liquid with the non-local boundary condition Eq.(\ref{PS_grid}) (see Appendix~\ref{ap:A} for details). We stress again that the resulting problem is not affected by the texture relief or the method used to find $P_{ij}^{x,y}$ due to a half-space liquid domain. From the liquid velocity field we can extract both the local slip length profile $b_{x,y}(y)$ (by using Eq.~\ref{localB}) and the effective slip tensor $b_{\rm eff}^{\|,\perp}$ (by averaging over texture period) as will be discussed below.

\section{Results and discussion.}

\subsection{Local slip length}
\begin{figure}[h]
\includegraphics[width=0.5\textwidth]{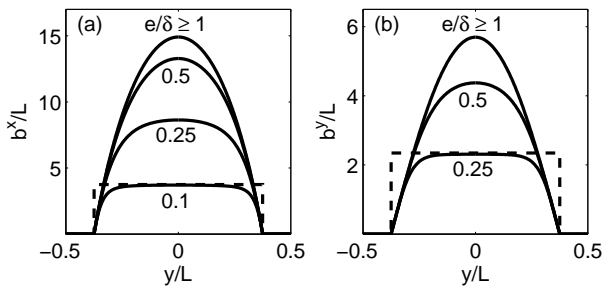}
\caption{Longitudinal (a) and transverse (b) local slip lengths computed with $\phi=0.75$, $\mu/\mu_g=50$. Dashed lines show the predictions of Eq.(\ref{bgas}).  }
\label{blocal}
\end{figure}

Figs.~\ref{blocal}(a) and (b) show profiles of the longitudinal, $b^x(y)$, and transverse, $b^y(y)$,  local slip lengths  at fixed groove width, $\delta/L=0.75$, and aspect ratio, $e/\delta$, varying from $0.1$ to infinity. The calculations are made using $\mu/\mu_g = 50$, which corresponds to a SH texture filled with gas.  It can be seen that for shallow grooves, $e/\delta\ll 1$, local slip lengths saturate to constant values predicted by Eq.(\ref{bgas}) at the central part of the gas sector, but $b^{x,y}(y)$ vanish at the edge of the groove. Thus the local slip profiles can be roughly approximated by a trapezoid~\cite{Zhou_etal:2013}.

For deeper grooves the local slip curves look more as parabolic. At $e/\delta\geq 1$ they converge to a single curve suggesting that $b^{x,y}(y)$ of deep grooves are controlled by the value of $\delta$ only, being independent on a texture depth. This result does not support Eq.(\ref{bgas}), which predicts that $b^{x,y}(y)$ are growing infinitely with $e$, and indicates that for large $e$ the dissipation at the edge of the grooves becomes crucial.

\begin{figure}[h]
\includegraphics[width=0.4\textwidth]{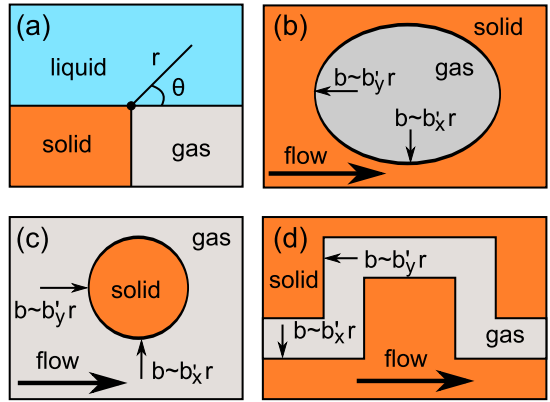}
\caption{ Polar coordinates (side view) used to evaluate the flow near the edge of the groove, (a) and illustration (top view) of the local slip length behavior at the edge of 2D pillars (b), hollows (c) and channels (d).}
\label{sketch2}
\end{figure}

Indeed, the data presented in Figs.~\ref{blocal}(a) and (b) suggest that near the edge of the groove $b^{x,y}$ always augment from zero by having the same slope (which has not been taken into account in recent work~\cite{shoenecker.c:2014}). This slope can be found by asymptotic analysis in the vicinity of the grooves edge. Motivated by an earlier single-phase analysis~\cite{wang2003,asmolov_etal:2013}, we can now construct the asymptotic solution for the two-phase flow near the edge by using polar coordinates $(r,\theta )$ (see Fig.~\ref{sketch2}(a) and Appendix ~\ref{ap:C} for details).

Close to the edge, when $r\ll 1$, the general solution of the Stokes equations implies a power-law dependence of velocities on the distance, $u \propto r^\lambda$: $u_{x}=r^{\lambda }a\sin \left( \lambda \theta \right)$, $u_{gx}=r^{\lambda }\left[ c\sin \left( \lambda \theta \right)
+h\cos \left( \lambda \theta \right)\right]$.
Similar arguments are valid for the transverse configuration. This yields a linear dependence for the slip lengths, $b^{x,y} = r \delta b'_{x,y}(\lambda)$. The exponent $\lambda$ can be found from the boundary conditions at the solid walls and the liauid/gas interface.

For large $\mu/\mu_g$ we obtain (see Appendix ~\ref{ap:C}):
\begin{equation}\label{slopes} b'_x\simeq 2\mu /\mu_g,\; b'_y\simeq \mu /(2\mu_g).\end{equation} The above expressions for $b_{x}^{\prime }$ and $b_{y}^{\prime }$ give upper and lower bounds on slopes among all textures, which are attained when the main flow is tangent or normal to the border of the gas area. Therefore, $b^{x}$ is constrained by $\delta\mu/\mu_g$, and $b^{y}$ by $\delta\mu/4\mu_g$ (see Fig.\ref{sketch2}(b-d)), so that the local slip profiles for arbitrary textures should be similar to shown in Fig.\ref{blocal}, although the absolute value of maximum might differ.
\begin{figure}[h]
\includegraphics[width=0.5\textwidth]{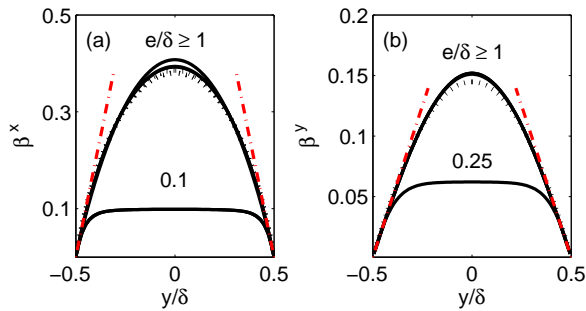}
\caption{Rescaled longitudinal (a) and transverse (b) local slip-length for deep and shallow grooves. Solid and dotted curves correspond to the Cassie, $\mu/\mu_g=50$, and Wenzel, $\mu/\mu_g=1$, states at $\phi=0.1, 0.5, 0.9$ (these curves coincide for shallow grooves and are nearly overlapping for deep grooves).  Dash-dotted lines show asymptotic solutions near the edges of the groove.}
\label{betalocal}
\end{figure}

It is natural now to propose a generalization of Eq.(\ref{bgas}) for a SH surface, where we scale with $\delta$ instead of $L$:
\begin{equation}
b^{x,y}(y) ={\delta} \dfrac{\mu}{\mu_g} \beta^{x,y}(e/\delta,y/\delta).\label{f}
\end{equation}
Here we ascribe rescaled dimensionless local slip lengths, $\beta^{x,y}$, which become linear in $e/\delta$ when $e/\delta$ is small,  and we recover Eq.(\ref{bgas}). At the other extreme, when $e/\delta$ is large,  $\beta^{x,y}$ saturate to provide an upper limit for local slip lengths.

To verify this Anzatz in Fig.~\ref{blocal} (a) and (b) we plot  $\beta^{x,y}$ as a function of $y/\delta$ at different $\phi$ and $e/\delta$. Here we use a viscosity ratio of the Cassie state as in Figs.~\ref{blocal}(a) and (b). Also included are results calculated for the Wenzel
state, and $\mu /\mu_g = 1$.  The results are somewhat remarkable. We see that for relatively deep grooves, $e/\delta \geq 1$, $\beta^{x,y}$ profiles computed for different  $\mu/\mu_g$, $e/\delta$ and even $\phi$, practically converge into a single curve~\cite{note1}, which coinsides with the numerical (but not semi-analytical) results reported before~\cite{shoenecker.c:2014}.

For  shallow grooves ($e/\delta\leq0.1$ for a longitudinal and $e/\delta\leq0.25$ for a transverse case) the $\beta^{x,y}$ profiles depend only on the depth of the groove, and can be approximated by trapezoids with the central region of a constant slip given by Eq.(\ref{bgas}), and linear edge regions where the local slip length is described by our asymptotic model. 

\subsection{Effective slip length}
\begin{figure}[h]
\includegraphics[width=0.5\textwidth]{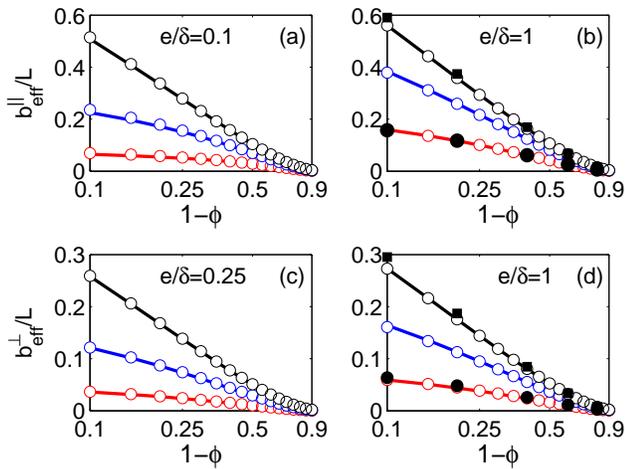}
\caption{Longitudinal (a,b) and transverse (c,d) effective slip lengths for textures with  shallow (a,c) and  deep (b,d) grooves. From top to bottom $\mu/\mu_g=50,5,1$. Exact theoretical results are shown by circles, analytical results [Eq.(\ref{belyaev})] with local slip given by Eq.(\ref{newb2}) are plotted by solid curves.  Filled squares show earlier data for perfect slip~\cite{philip.jr:1972,lauga.e:2003}, filled circles show earlier data for the Wentzel state~\cite{wang2003}. }
\label{figbeff}\end{figure}

We finally turn to the effective slip lengths, which can be found by averaging the obtained numerical solution for longitudinal and transverse directions:
\begin{equation}
 b_{\mathrm{eff}}^{\|,\perp}=\left.\dfrac{\langle u_{x,y}\rangle}{\langle \partial_z u_{x,y}\rangle}\right|_{z=0}.
\end{equation}
  The calculations are made using the viscosity ratio of the Cassie and Wenzel states. For completeness we include the data for $\mu/\mu_g=5$, which correspond to oil-impregnated textures. Fig.~\ref{figbeff} shows longitudinal (a,b) and transverse (c,d) effective slip lengths as a function of solid fraction, $1-\phi,$ for shallow (a,c) and deep (b,d) grooves.
The eigenvalues of the effective lengths of a striped surface with a \emph{piecewise constant} local slip, $b_c^{x,y}$, have been calculated analytically~\cite{belyaev.av:2010a}:

\begin{equation}
\begin{array}{ll}
b_\mathrm{eff}^{\|}\simeq\dfrac{L}{\pi}\dfrac{\ln\left[\sec\left(\frac{\pi\phi}{2}\right)\right]}{1+\dfrac{L}{\pi b_c^x}\ln\left[\sec\left(\frac{\pi\phi}{2}\right)+\tan\left(\frac{\pi\phi}{2}\right)\right]},
\\\label{belyaev}
b_\mathrm{eff}^{\perp}\simeq\dfrac{L}{2\pi}\dfrac{\ln\left[\sec\left(\frac{\pi\phi}{2}\right)\right]}{1+\dfrac{L}{2\pi b_c^y}\ln\left[\sec\left(\frac{\pi\phi}{2}\right)+\tan\left(\frac{\pi\phi}{2}\right)\right]}.
\end{array}\end{equation}

\begin{figure}[h]
\includegraphics[width=0.24\textwidth]{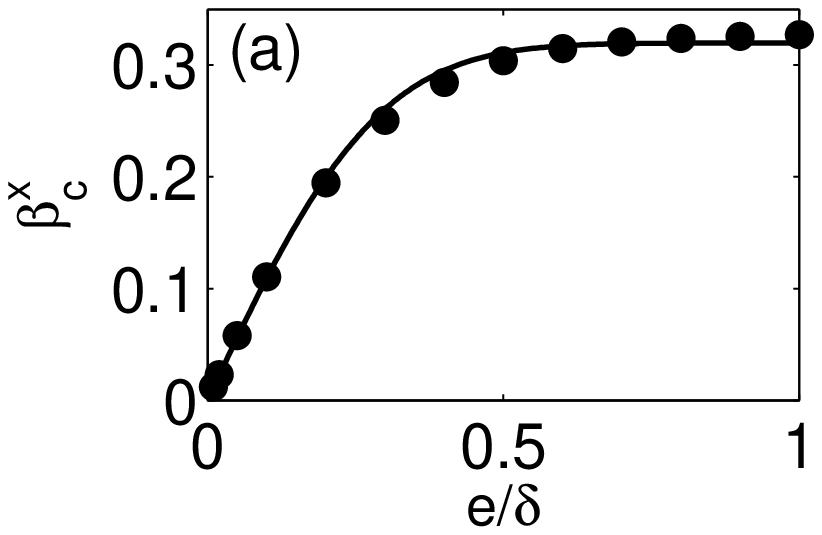}\includegraphics[width=0.24\textwidth]{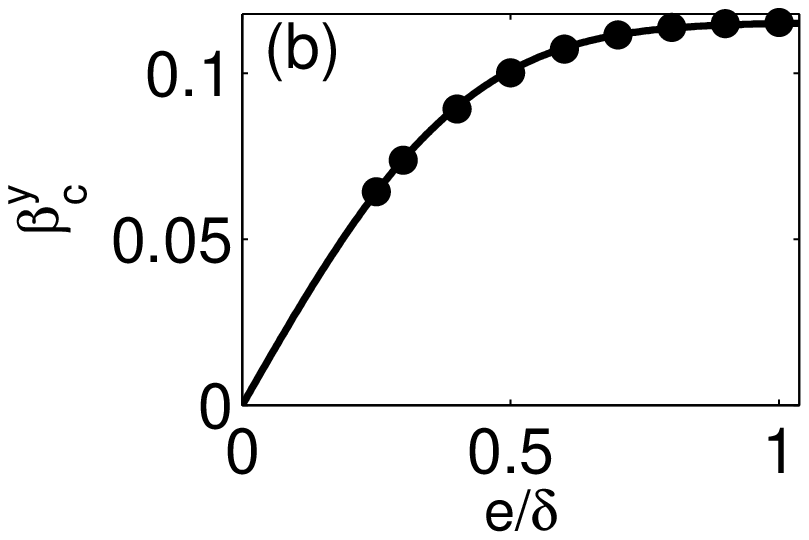}
\caption{Apparent local slip as a function of the depth of the groove for longitudinal (a) and transverse (b) directions. Symbols show the values obtained using Eqs.(\ref{belyaev}), solid curves show predictions of Eqs.(\ref{fits}).}
\label{apparent}\end{figure}

Let us now try to define \emph{apparent} constant local slip lengths at the gas sectors. Eq.(\ref{f}) suggests the following definition
\begin{equation}
b_{c}^{x,y}=\delta \dfrac{\mu }{\mu _{g}}\beta _c^{x,y},  \label{newb2}
\end{equation}%
where dimensionless slip lengths, $\beta _{c}^{x,y},$ depend only on the
aspect ratio of the texture, $e/\delta $. We fitted our theoretical results for $\mu/\mu_g = 5$ to Eq.(\ref{belyaev}) taking
$\beta^{x,y}_c$ as a fitting parameter. The obtained values are surprisingly well described by simple functions

\begin{equation}\label{fits}
\beta_c^{x}\simeq\dfrac{\mathrm{erf} {(q}_{x}{e/\delta })}{q_{x}}, \;
\beta _c^{y}\simeq\dfrac{\mathrm{erf} {(q}_{y}{e/\delta })}{4q_y},\end{equation}
with ${q}_{x}\simeq3.1$ ${q}_{y}\simeq2.17$. These functions saturate to $\beta _{c}^{x} \simeq 0.32$ and $\beta _{c}^{y} \simeq 0.12$ already at $e/\delta\geq 1$, by imposing constraints on the attainable $b_{c}^{x,y}$ (see Fig.~\ref{apparent}).

Assuming $\beta_{c}^{x,y}$ found for $\mu/\mu_g = 5$ are universal, we can then use Eq.({\ref{belyaev}}) to calculate the effective slip lengths for $\mu/\mu_g = 1$ and 50. The results are included in Fig.\ref{figbeff}.
A general
conclusion is that the predictions of Eq.({\ref{belyaev}}) with the local
slip defined by Eq.({\ref{newb2}}) are in excellent
agreement with exact theoretical results in the whole range of parameters,
$0.1\leq \phi \leq 0.9$ and $\mu /\mu _{g}\geq 1$, confirming the universality of $\beta_{c}^{x,y}$. Note that included in Fig.%
\ref{figbeff} effective slip lengths for perfect-slip stripes~\cite%
{philip.jr:1972,lauga.e:2003}, practically coincide with our results for
$\mu /\mu _{g}=50$. We can then conclude that SH surfaces in the Cassie state
provide the very general upper bound for effective slip of textured
surfaces, valid for whatever large viscosity contrast (e.g. polymer melts~%
\cite{degennes:1979}). Finally, we observe an excellent agreement of our
results for $\mu/\mu_g=1$ with earlier data even for the Wenzel state obtained by using a completely different
approach~\cite{wang2003}.

Now, we recall that for pillars in the low $\phi$ limit, $\delta\simeq L$, the average local slip was shown to scale as~\cite{ybert.c:2007}:
\begin{equation}\label{bocquet}
   b^{x,y}_a\simeq L\frac{\mu }{\mu _{g}}\beta \tanh \left( \frac{e}{L \beta}\right)
\end{equation}
with $\beta(\phi)$. For deep dilute pillars Eq.(\ref{bocquet}) transforms to $\displaystyle b^{x,y}_a\simeq L \frac{\mu }{\mu _{g}} \beta$ (cf. Eq.(\ref{newb2})), and for pillars with $\phi =0.9$
we evaluate $\beta\simeq 0.2$~\cite{ybert.c:2007}. This value is close to the exact ones found here for deep rectangular grooves, $\beta_{c}^{x,y}$, so our theory
provides a good sense of the possible local slip of 2D texture.

\section{ Conclusion.}
We have proposed an operator method, which allowed us the mapping of the flow in the gas  subphase to a local slip boundary condition at the gas area of SH surfaces. The determined slip length is shown to be a unique function of the viscosity
contrast and  topography of the underlying texture. Our main results, Eqs.({\ref{f}}) and ({\ref{newb2}}), can be thus viewed as a general `gas cushion' model for textured surfaces, which transforms to the standard model, Eq.({\ref{bgas}}), in case of shallow textures. We have proven that besides Cassie surfaces our approach is valid for Wenzel textures, as well as rough surfaces impregnated by a `lubricant' with
lower viscosity.

We checked the validity of our approach by studying a flow past canonical rectangular grooves, but our strategy
can be immediately applied for 1D textures with different cross-sections or extended to more complex 2D textures.
These textures include various pillars, and holes and lamellae of a complex shape.
Thus, our results may guide the design of textured surfaces with superlubricating potential in microfluidic devices, tribology, polymer science, and more. Another fruitful direction could be to apply our method to calculations of an electro-osmotic~\cite{belyaev.av:2011a,Squires08,bahga:2010} and diffusio-osmotic ~\cite{Huang08} flow past textured surfaces.

\appendix
\section{Numerical solution in liquid}\label{ap:A}

The non-local boundary condition given by Eq.~(\ref{PS_grid}) is easy to implement with the Fourier method. For simplicity we consider only the longitudinal flow here, the procedure for the transverse flow is similar. We present the flow over the superhydrophobic surface as a sum of the undisturbed flow and the correction due to the slip:
$$u_x(y,z)=u_0(z)+u_1(y,z),$$
where $u_0(z)=Gz$ is a simple shear flow and $G$ is the undisturbed shear rate.
The correction is sought in the form of a cosine series, since the solution is symmetric in $y$:
$$u_1(y,z)=c_0+\sum\limits_{m=1}^{\infty}c_m \cos(k_m y)\exp(-k_mz),$$
where $k_m=2\pi m/L$.

Boundary conditions for the correction read:
\begin{equation}
\begin{array}{ll}
\dfrac{\partial u_1}{\partial z}-\mathbf P^x[u_1]=-G, &0<y<\delta/2,\\
u_1=0, &\delta/2<y<L/2.\\
\end{array}
\label{PS}\end{equation}
(due to the symmetry of the problem, it is sufficient to consider only a half of the period). We take the same spatial grid that is used for the DtN matrix: $y_i=\delta\eta_i$  with $N$ nodes over the groove
$$y_i=\delta(i-1)/2N, \quad i=1\dots N$$ and add $M$ nodes at the boundary in contact with solid: $$y_i=(L-\delta)(i-1)/[2(M-1)]+\delta/2, \quad i=N+1\dots N+M.$$
We cut the series to $N_f=N+M$ terms and obtain the following linear system:
\begin{equation}\begin{array}{ll}
\sum\limits_{m=1}^{N_f}D_{ik}c_m=-G,  &i=1\dots N,\\
c_0+\sum\limits_{m=1}^{N_f}c_m\cos(k_my_i)=0,  &i=N+1\dots N_f,\\
\label{discrete}
\end{array}\end{equation}
$$D_{in}=k_n\cos(k_n y_{i})-\sum\limits_{j=1}^{N} P^x_{ij}\cos(k_ny_{j}).$$
Once the matrix $P^x_{ij}$ is known, the linear system (\ref{discrete}) can be solved to find the correction, and the complete velocity field in liquid can be calculated. A similar procedure can be applied to the transverse flow (see \cite{nizkaya.tv:2013} for reference).

\section{Dirichlet-to-Neumann matrix for a rectangular groove}\label{ap:B}
For rectangular geometries the Fourier method can be implemented to calculate the DtN matrices.

\textbf{Longitudinal configuration.} We introduce the non-dimensional variables in the gas domain, $\eta=y/\delta$ and $\zeta=z/\delta$. Following \cite{ng:2010}, we seek for the solution of the Laplace equation in gas in the form:
\begin{equation}
u_{gx}(\eta,\zeta)=\sum\limits_{m=1}^{\infty}c_m\sinh[k^*_m(\zeta+d)]\cos(k^*_m \eta),
\label{cos}\end{equation}
where $k_m^*=(2m-1)\pi$, $d=e/\delta$. Each term in this series is a partial solution of the Laplace equation, which is symmetric in $\eta$ and satisfies the no-slip boundary conditions at the side walls, $\eta=\pm1/2$, and at the bottom wall of the groove, $\zeta=-d$, (see Fig.~\ref{sketch}).

The velocity and its normal derivative at the liquid/gas interface are
\begin{equation}\begin{array}{ll}
\zeta=0: \; u_{gx}=\sum\limits_{m=1}^{\infty}c_m\sinh(k^*_md)\cos(k^*_m \eta),\\
\dfrac{\partial u_{gx}}{\partial \zeta}=\sum\limits_{m=1}^{\infty}k^*_mc_m\cosh(k^*_md)\cos(k^*_m \eta).\end{array}\label{ududn}\end{equation}
The relation between them can be obtained from (\ref{ududn}):
$$\dfrac{\partial u_{gx}}{\partial \zeta}=\sum\limits_{m=1}^{\infty}\left(\Pi_{mn}u^*_n\right)\cos(k^*_m \eta),$$
$$u_n^*=c_n\sinh(k^*_nd),$$
$$\Pi_{ml}=k^*_m\coth(k_m^*e/\delta)\delta_{ml},\; m,l=1..\infty.$$
Here $\Pi_{ml}$ is the representation of the  DtN operator in the Fourier space.

The last step is to transform this operator into physical space, so that it can be applied as a boundary condition. To do so, we introduce a spatial grid $\eta_i=(i-1)/(2N),\;i=1..N$ at the liquid/gas interface (due to the symmetry it is sufficient to consider only a half of the period) and cut the cosine series (\ref{cos}) to $N$ terms accordingly. Considering Eq. (\ref{ududn}) at each grid node we have:
$$u_{gx}(\eta_i,0)=\sum_{m=1}^{N}F_{im}u^*_m,$$
and hence,
$$u^*_m=\sum_{i=1}^{N}F^{-1}_{mi}u_{gx}(\eta_i,0),
$$
where $F_{im}=\cos(k^*_m \eta_i)$ is the collocation matrix.

The normal derivative of the velocity (using dimensional variables $y,z$) reads:
$$\dfrac{\partial u_{gx}(\eta_i,0)}{\partial z}=\sum_{j=1}^{N}P^x_{ij}u_{gx}(\eta_j,0),$$
where $$P^x_{ij}=\dfrac{1}{\delta}\left( F_{im} \Pi_{ml}  F^{-1}_{lj}\right),$$
is the DtN matrix for the longitudinal flow.

\textbf{Transverse configuration}. We assume that the liquid/gas interface is flat, so that $u_{gz}=0$ at $\zeta=0$. The symmetry condition implies that $u_{gy}$ is symmetric in $\eta$ while $u_{gz}$ is anti-symmetric.
We represent the solution in gas in the following form~\cite{ng:2010}:
\begin{equation*}
\begin{array}{ll}
u_{gy}(\eta,\zeta)=\sum\limits_{n=1}^{\infty}\dfrac{\cos(k^*_n\eta)}{k^*_n\cosh(k^*_n d)}\left[A_nK'_n(\zeta) + B_nG'_n(\zeta)\right]\\
\displaystyle+\sum\limits_{n=1}^{\infty}\cos(\beta_n\zeta)D_nH_n(\eta),
\end{array}
\end{equation*}
\begin{equation*}
\begin{array}{ll}
u_{gz}(\eta,\zeta)=\sum\limits_{n=1}^{\infty}\dfrac{\sin(k^*_n\eta)}{\cosh(k^*_n d)}\left[A_nK_n(\zeta) + B_nG_n(\zeta)\right]\\
\displaystyle-\sum\limits_{n=1}^{\infty}\sin(\beta_n\zeta)\dfrac{D_n}{\beta_n}H'_n(\eta),
\end{array}
\end{equation*}
\begin{eqnarray*}
K_n &=& \sinh(k^*_n\zeta) - \zeta \exp[-k^*_n(\zeta + d)]\sinh(k^*_n d)/d,\\
G_n &=& k^*_n \zeta \sinh[k^*_n(\zeta + d)],\\
H_n &=& 2\exp(-\beta_n/2)\left[\cosh(\beta_n\eta) - 2\eta\coth(\beta_n/2)\sinh(\beta_n\eta)\right].
\end{eqnarray*}
where $k^*_n=(2n-1)\pi$ and $\beta_n=n\pi d$; $A_n,\;B_n$ and $D_n$ are the unknown coefficients.
The conditions of non-permeability at the side walls, $u_{gy}=0$ at $\eta=\pm1/2$, and at the bottom wall and at the interface, $u_{gz}=0$ at $\zeta=0,-d$, are satisfied automatically. The no-slip boundary conditions at the walls of the groove ($u_{gz}=0$ at $\eta=\pm1/2$ and $u_{gy}=0$ at $\zeta=-d$) and the continuity condition at the interface ($u_{gy}=v^*$ at $\zeta=0$) have to be satisfied by a proper choice of the coefficients $A_n, B_n, D_n$. To do so, we cut the series to $N$ terms and introduce a grid covering the walls of the groove and the interface and containing $3N$ nodes ($N$ at each wall/interface). Calculating the tangential velocity at each point of the groove, we obtain a system of $3N$ linear equations for a $3N$-component vector $Z_k=\{A_1,\dots A_N, B_1\dots, B_N, D_1,\dots D_n\}$. The right-hand sides of the equations are equal to zero at groove's walls (no-slip) and to liquid velocity at the interface, $u_{gy}(0,\eta_i)= v^*(\eta_i)$. The solution satisfying the no-slip boundary conditions and taking the prescribed values at the interface can be expressed in a matrix form:
$$Z_k=\sum\limits_{j=0}^{N}M_{kj}v^*(\eta_j),$$
where $v^*(\eta_j)$ is a $N$-component vector of  velocity at the interface grid points and $M_{kj}$ is a $3N\times N$ matrix.
Then the normal derivative at the interface can be expressed in the following way:
\begin{eqnarray*}
\dfrac{\partial v_{g}(\eta _{j},0)}{\partial \zeta } &=&\sum%
\limits_{n=1}^{N}2\left[ \dfrac{A_{n}}{d}\exp (-k^*_nd)\tanh (k^*_n d)\right.  \\
& +&B_{n}k^*_n\Big]\cos (k^*_n\eta _{j})=Q_{ik}Z_{k},
\end{eqnarray*}
where $Q_{ik}$ is $N\times 3N$ matrix.

Back in dimensional variables $y,z$, we obtain the following representation for the $N\times N$ DtN matrix :
$$P^y_{ij}=\delta^{-1}Q_{ik}M_{kj}.$$
\
\section{Asymptotic solution near the edge of the grooves}
\label{ap:C}

Here we obtain a solution in the vicinity of
the groove corner by using polar coordinates $(r,\theta )$ \cite%
{wang2003, Zhou_etal:2013}, with the origin at $\left( y_{c},z_{c}\right)
=\left(-\delta/2 ,0\right) $, so that $y=y_{c}+r\delta\cos \theta ,$ $z=r\delta\sin
\theta $ (see Fig.~\ref{sketch2}(a)). Similar approach has been applied earlier
for single-phase flows to describe singularities near sharp
corners. For the flow over a surface with rectangular grooves, the shear stress has found to be singular{, i.e.,} proportional to $r^{-1/3}$ for longitudinal and {to} $%
r^{-0.455}$ for transverse configurations~\cite{wang2003}. The edge between different slipping flat interfaces has also been considered, with alternating
no-slip and slip stripes \cite%
{asmolov:2012, asmolov_etal:2013}, trapezoidal and triangular profiles of the local slip $b(y)$ \cite%
{Zhou_etal:2013}.

For the two-phase flow near the corner of a groove with flat interface, the liquid/solid, liquid/gas and gas/solid interfaces correspond to $\theta =0,\ \theta =\pi $ and $\theta =3\pi /2$ (if the wall of the groove is vertical). A general solution of the dimensionless Laplace equation is a power dependence on the distance $r$:%
\begin{eqnarray}
r\ll 1: \; u_{x}&=&r^{\lambda }\left[ a\sin \left( \lambda \theta \right)
+g\cos \left( \lambda \theta \right) \right],
\label{u_g1}\\
u_{gx}&=&r^{\lambda }\left[ c\sin \left( \lambda \theta \right)
+h\cos \left( \lambda \theta \right) \right],
\label{u_g2}
\end{eqnarray}
where $a$, $g$, $c$ and $h$ are constants which may be found by matching (%
\ref{u_g1}), (\ref{u_g2}) with the flow at $r\sim 1.$ However, the exponent $\lambda $ can be obtained solely from the boundary conditions.

The no-slip boundary condition for liquid phase at $\theta =0$, the no-slip boundary condition for gas phase at $\theta =3\pi /2$ and the coupling conditions at the gas/liquid interface, Eq.(\ref{BCcon}), lead to a linear system on $a,\ g,\ c,\ h $ which yields the following equation on $\lambda$:
\begin{equation}
\tan \left( \frac{3\lambda \pi}{2}\right) =\frac{\left( \mu-\mu_g\right)
\tan \left( \lambda \pi \right) }{\mu+\mu_g\tan^2\left( \lambda \pi
\right) }.  \label{lam1}
\end{equation}%
Thus the exponent depends on the viscosity ratio $\mu/\mu_g$ only. Previous
analytical solutions, $\lambda =2/3,\ $ for a single-phase rectangular hydrophilic groove with $%
\mu/\mu_g=1\ $~\cite{wang2003}, and $\lambda _{\mathrm{ideal}}=1/2$ for a flat shear-free interface with $\mu/\mu_g=\infty \ $\cite{philip.jr:1972,asmolov:2012} satisfy the equation obtained. When the viscosity ratio is large, as for liquid/gas
case, we construct an asymptotic solution of (\ref{lam1}) in terms of series
in $\mu_g/\mu\ll 1:$
\begin{equation}
\lambda =\frac{1}{2}-\frac{\mu_g}{\mu\pi }+O\left(\dfrac{\mu_g^2}{\mu^2}%
\right).  \label{exp}
\end{equation}%
The asymptotic solution (\ref{exp}) is close to that for alternating no-slip
and perfect-slip stripes with $\lambda _{\mathrm{ideal}}=1/2$. The local
slip length near the edge can be defined as%
\[
r\ll 1:\quad b_x(r)=\frac{u_{x}(r, \pi)}{\partial _\theta u_x(r, \pi )}
\simeq 2\frac{\mu }{\mu_g}r\delta.
\]%
Therefore, $b_x$ is linear in the distance $r$ from the corner at the
liquid/gas interface. The slope of the dependence is large, of order of $%
\mu/\mu_g$.

For the flow transverse to the grooves, we
represent the solution in liquid in terms of a streamfunction $\psi $ which
satisfies a biharmonic equation $\Delta ^{2}\psi =0$. A general solution can
be presented in the form~\cite{wang2003}:
\begin{equation}
\begin{array}{ll}
\psi (r,\theta ) & =r^{\lambda }\left[ a\sin (\lambda \theta )+g\sin
((\lambda -2)\theta )\right.  \\
& +\left. c\cos (\lambda \theta )+h\cos ((\lambda -2)\theta )\right]
.%
\end{array}
\label{tr1}
\end{equation}%
The radial and the angular components of the liquid velocity are
\begin{equation}
u_{r}(r,\theta )=\dfrac{\partial _{\theta }\psi }{r},\;u_{\theta }(r,\theta
)=-\partial _{r}\psi .  \label{tr2}
\end{equation}%
Equations similar to (\ref{tr1}), (\ref{tr2}) can be also written for the gas
streamfunction $\psi _{g}$ and velocity components $u_{gr},$ $u_{g\theta }.$
We apply the no-slip boundary conditions at $\theta =0$ and $\theta =3\pi /2$
and the continuity conditions at the gas/liquid interface, $\theta =\pi $,
and, similar to (\ref{lam1}), we obtain the equation governing $\lambda $:
\begin{equation}
\begin{array}{lll}
\dfrac{\tan (\lambda \pi )}{2(\lambda -1)}=\dfrac{\mu }{\mu _{g}}\dfrac{%
2\lambda ^{2}-4\lambda +1+\cos (\lambda \pi )}{4(1-\lambda )\cos ^{2}{\left( \lambda \pi /2\right) }}, &  &
\end{array}
\label{lam2}
\end{equation}%
For a shear-free interface, $\mu /\mu _{g}=\infty ,$ we have from (\ref{lam2})
$\lambda _{\mathrm{ideal}}=1/2$. Therefore, for large $\mu /\mu _{g}$ we can
again construct an asymptotic solution of (\ref{lam2}). The local slip
length, to the first order in $\mu _{g}/\mu $, reads
\[
r\ll 1:\quad b_{y}(r)=\frac{u_{r}\left( r,\pi \right) }{\partial _{\theta
}u_{r}\left( r,\pi \right) }\simeq \dfrac{1}{2}\dfrac{\mu }{\mu _{g}}r\delta.
\]

{\footnotesize{\bibliography{prl} 

\begin{thebibliography}{39}
\expandafter\ifx\csname natexlab\endcsname\relax\def\natexlab#1{#1}\fi
\expandafter\ifx\csname bibnamefont\endcsname\relax
  \def\bibnamefont#1{#1}\fi
\expandafter\ifx\csname bibfnamefont\endcsname\relax
  \def\bibfnamefont#1{#1}\fi
\expandafter\ifx\csname citenamefont\endcsname\relax
  \def\citenamefont#1{#1}\fi
\expandafter\ifx\csname url\endcsname\relax
  \def\url#1{\texttt{#1}}\fi
\expandafter\ifx\csname urlprefix\endcsname\relax\def\urlprefix{URL }\fi
\providecommand{\bibinfo}[2]{#2}
\providecommand{\eprint}[2][]{\url{#2}}

\bibitem[{\citenamefont{Quere}(2008)}]{quere.d:2008}
\bibinfo{author}{\bibfnamefont{D.}~\bibnamefont{Quere}},
  \bibinfo{journal}{Annu. Rev. Mater. Res.} \textbf{\bibinfo{volume}{38}},
  \bibinfo{pages}{71} (\bibinfo{year}{2008}).

\bibitem[{\citenamefont{{Bocquet} and Barrat}(2007)}]{bocquet2007}
\bibinfo{author}{\bibfnamefont{L.}~\bibnamefont{{Bocquet}}} \bibnamefont{and}
  \bibinfo{author}{\bibfnamefont{J.~L.} \bibnamefont{Barrat}},
  \bibinfo{journal}{Soft Matter} \textbf{\bibinfo{volume}{3}},
  \bibinfo{pages}{685} (\bibinfo{year}{2007}).

\bibitem[{\citenamefont{Rothstein}(2010)}]{rothstein.jp:2010}
\bibinfo{author}{\bibfnamefont{J.~P.} \bibnamefont{Rothstein}},
  \bibinfo{journal}{Annu. Rev. Fluid Mech.} \textbf{\bibinfo{volume}{42}},
  \bibinfo{pages}{89} (\bibinfo{year}{2010}).

\bibitem[{\citenamefont{{Vinogradova} and {Dubov}}(2012)}]{vinogradova.oi:2012}
\bibinfo{author}{\bibfnamefont{O.~I.} \bibnamefont{{Vinogradova}}}
  \bibnamefont{and} \bibinfo{author}{\bibfnamefont{A.~L.}
  \bibnamefont{{Dubov}}}, \bibinfo{journal}{Mendeleev Commun.}
  \textbf{\bibinfo{volume}{19}}, \bibinfo{pages}{229} (\bibinfo{year}{2012}).

\bibitem[{\citenamefont{Voronov et~al.}(2008)\citenamefont{Voronov,
  Papavassiliou, and Lee}}]{voronov.rs:2008}
\bibinfo{author}{\bibfnamefont{R.~S.} \bibnamefont{Voronov}},
  \bibinfo{author}{\bibfnamefont{D.~V.} \bibnamefont{Papavassiliou}},
  \bibnamefont{and} \bibinfo{author}{\bibfnamefont{L.~L.} \bibnamefont{Lee}},
  \bibinfo{journal}{Ind. Eng. Chem. Res.} \textbf{\bibinfo{volume}{47}},
  \bibinfo{pages}{2455} (\bibinfo{year}{2008}).

\bibitem[{\citenamefont{Vinogradova and Belyaev}(2011)}]{vinogradova.oi:2011}
\bibinfo{author}{\bibfnamefont{O.~I.} \bibnamefont{Vinogradova}}
  \bibnamefont{and} \bibinfo{author}{\bibfnamefont{A.~V.}
  \bibnamefont{Belyaev}}, \bibinfo{journal}{J. Phys.: Condens. Matter}
  \textbf{\bibinfo{volume}{23}}, \bibinfo{pages}{184104}
  (\bibinfo{year}{2011}).

\bibitem[{\citenamefont{Kamrin et~al.}(2010)\citenamefont{Kamrin, Bazant, and
  Stone}}]{Kamrin_etal:2010}
\bibinfo{author}{\bibfnamefont{K.}~\bibnamefont{Kamrin}},
  \bibinfo{author}{\bibfnamefont{M.~Z.} \bibnamefont{Bazant}},
  \bibnamefont{and} \bibinfo{author}{\bibfnamefont{H.~A.} \bibnamefont{Stone}},
  \bibinfo{journal}{J.~Fluid Mech.} \textbf{\bibinfo{volume}{658}},
  \bibinfo{pages}{409} (\bibinfo{year}{2010}).

\bibitem[{\citenamefont{{Stone} et~al.}(2004)\citenamefont{{Stone}, {Stroock},
  and {Ajdari}}}]{stone2004}
\bibinfo{author}{\bibfnamefont{H.~A.} \bibnamefont{{Stone}}},
  \bibinfo{author}{\bibfnamefont{A.~D.} \bibnamefont{{Stroock}}},
  \bibnamefont{and} \bibinfo{author}{\bibfnamefont{A.}~\bibnamefont{{Ajdari}}},
  \bibinfo{journal}{Annual Review of Fluid Mechanics}
  \textbf{\bibinfo{volume}{36}}, \bibinfo{pages}{381} (\bibinfo{year}{2004}).

\bibitem[{\citenamefont{Bazant and Vinogradova}(2008)}]{Bazant08}
\bibinfo{author}{\bibfnamefont{M.~Z.} \bibnamefont{Bazant}} \bibnamefont{and}
  \bibinfo{author}{\bibfnamefont{O.~I.} \bibnamefont{Vinogradova}},
  \bibinfo{journal}{J.~Fluid Mech.} \textbf{\bibinfo{volume}{613}},
  \bibinfo{pages}{125} (\bibinfo{year}{2008}).

\bibitem[{\citenamefont{Philip}(1972)}]{philip.jr:1972}
\bibinfo{author}{\bibfnamefont{J.~R.} \bibnamefont{Philip}},
  \bibinfo{journal}{J. Appl. Math. Phys.} \textbf{\bibinfo{volume}{23}},
  \bibinfo{pages}{353} (\bibinfo{year}{1972}).

\bibitem[{\citenamefont{Priezjev et~al.}(2005)\citenamefont{Priezjev, Darhuber,
  and Troian}}]{priezjev.nv:2005}
\bibinfo{author}{\bibfnamefont{N.~V.} \bibnamefont{Priezjev}},
  \bibinfo{author}{\bibfnamefont{A.~A.} \bibnamefont{Darhuber}},
  \bibnamefont{and} \bibinfo{author}{\bibfnamefont{S.~M.}
  \bibnamefont{Troian}}, \bibinfo{journal}{Phys. Rev. E}
  \textbf{\bibinfo{volume}{71}}, \bibinfo{pages}{041608}
  (\bibinfo{year}{2005}).

\bibitem[{\citenamefont{Lauga and Stone}(2003)}]{lauga.e:2003}
\bibinfo{author}{\bibfnamefont{E.}~\bibnamefont{Lauga}} \bibnamefont{and}
  \bibinfo{author}{\bibfnamefont{H.~A.} \bibnamefont{Stone}},
  \bibinfo{journal}{J.~Fluid Mech.} \textbf{\bibinfo{volume}{489}},
  \bibinfo{pages}{55} (\bibinfo{year}{2003}).

\bibitem[{\citenamefont{Maynes et~al.}(2007)\citenamefont{Maynes, Jeffs,
  Woolford, and Webb}}]{maynes2007}
\bibinfo{author}{\bibfnamefont{D.}~\bibnamefont{Maynes}},
  \bibinfo{author}{\bibfnamefont{K.}~\bibnamefont{Jeffs}},
  \bibinfo{author}{\bibfnamefont{B.}~\bibnamefont{Woolford}}, \bibnamefont{and}
  \bibinfo{author}{\bibfnamefont{B.~W.} \bibnamefont{Webb}},
  \bibinfo{journal}{Phys. Fluids} \textbf{\bibinfo{volume}{19}},
  \bibinfo{pages}{093603} (\bibinfo{year}{2007}).

\bibitem[{\citenamefont{Ng et~al.}(2010)\citenamefont{Ng, Chu, and
  Wang}}]{ng:2010}
\bibinfo{author}{\bibfnamefont{C.}~\bibnamefont{Ng}},
  \bibinfo{author}{\bibfnamefont{H.}~\bibnamefont{Chu}}, \bibnamefont{and}
  \bibinfo{author}{\bibfnamefont{C.}~\bibnamefont{Wang}},
  \bibinfo{journal}{Phys. Fluids} \textbf{\bibinfo{volume}{22}},
  \bibinfo{pages}{102002} (\bibinfo{year}{2010}).

\bibitem[{\citenamefont{Sch{\"o}necker
  et~al.}(2014)\citenamefont{Sch{\"o}necker, Baier, and
  Hardt}}]{shoenecker.c:2014}
\bibinfo{author}{\bibfnamefont{C.}~\bibnamefont{Sch{\"o}necker}},
  \bibinfo{author}{\bibfnamefont{T.}~\bibnamefont{Baier}}, \bibnamefont{and}
  \bibinfo{author}{\bibfnamefont{S.}~\bibnamefont{Hardt}}, \bibinfo{journal}{J.
  Fluid Mech.} \textbf{\bibinfo{volume}{740}}, \bibinfo{pages}{168}
  (\bibinfo{year}{2014}).

\bibitem[{\citenamefont{Belyaev and Vinogradova}(2010)}]{belyaev.av:2010a}
\bibinfo{author}{\bibfnamefont{A.~V.} \bibnamefont{Belyaev}} \bibnamefont{and}
  \bibinfo{author}{\bibfnamefont{O.~I.} \bibnamefont{Vinogradova}},
  \bibinfo{journal}{J.~Fluid Mech.} \textbf{\bibinfo{volume}{652}},
  \bibinfo{pages}{489} (\bibinfo{year}{2010}).

\bibitem[{\citenamefont{Vinogradova}(1995)}]{vinogradova.oi:1995a}
\bibinfo{author}{\bibfnamefont{O.~I.} \bibnamefont{Vinogradova}},
  \bibinfo{journal}{Langmuir} \textbf{\bibinfo{volume}{11}},
  \bibinfo{pages}{2213} (\bibinfo{year}{1995}).

\bibitem[{\citenamefont{Nizkaya et~al.}(2013)\citenamefont{Nizkaya, Asmolov,
  and Vinogradova}}]{nizkaya.tv:2013}
\bibinfo{author}{\bibfnamefont{T.~V.} \bibnamefont{Nizkaya}},
  \bibinfo{author}{\bibfnamefont{E.~S.} \bibnamefont{Asmolov}},
  \bibnamefont{and} \bibinfo{author}{\bibfnamefont{O.~I.}
  \bibnamefont{Vinogradova}}, \bibinfo{journal}{Soft Matter}
  \textbf{\bibinfo{volume}{9}}, \bibinfo{pages}{11671} (\bibinfo{year}{2013}).

\bibitem[{\citenamefont{Sbragaglia and Prosperetti}(2007)}]{sbragaglia.m:2007}
\bibinfo{author}{\bibfnamefont{M.}~\bibnamefont{Sbragaglia}} \bibnamefont{and}
  \bibinfo{author}{\bibfnamefont{A.}~\bibnamefont{Prosperetti}},
  \bibinfo{journal}{Phys. Fluids} \textbf{\bibinfo{volume}{19}},
  \bibinfo{pages}{043603} (\bibinfo{year}{2007}).

\bibitem[{\citenamefont{Hyv\"{a}luoma and Harting}(2008)}]{harting.j:2008}
\bibinfo{author}{\bibfnamefont{J.}~\bibnamefont{Hyv\"{a}luoma}}
  \bibnamefont{and} \bibinfo{author}{\bibfnamefont{J.}~\bibnamefont{Harting}},
  \bibinfo{journal}{Phys. Rev. Lett.} \textbf{\bibinfo{volume}{100}},
  \bibinfo{pages}{246001} (\bibinfo{year}{2008}).

\bibitem[{\citenamefont{Ybert et~al.}(2007)\citenamefont{Ybert, Barentin,
  Cottin-Bizonne, Joseph, and Bocquet}}]{ybert.c:2007}
\bibinfo{author}{\bibfnamefont{C.}~\bibnamefont{Ybert}},
  \bibinfo{author}{\bibfnamefont{C.}~\bibnamefont{Barentin}},
  \bibinfo{author}{\bibfnamefont{C.}~\bibnamefont{Cottin-Bizonne}},
  \bibinfo{author}{\bibfnamefont{P.}~\bibnamefont{Joseph}}, \bibnamefont{and}
  \bibinfo{author}{\bibfnamefont{L.}~\bibnamefont{Bocquet}},
  \bibinfo{journal}{Phys. Fluids} \textbf{\bibinfo{volume}{19}},
  \bibinfo{pages}{123601} (\bibinfo{year}{2007}).

\bibitem[{\citenamefont{Karatay et~al.}(2013)\citenamefont{Karatay, Haase,
  Visser, Sun, Lohse, Tsai, and Lammertink}}]{karatay.e:2013}
\bibinfo{author}{\bibfnamefont{E.}~\bibnamefont{Karatay}},
  \bibinfo{author}{\bibfnamefont{A.~S.} \bibnamefont{Haase}},
  \bibinfo{author}{\bibfnamefont{C.~W.} \bibnamefont{Visser}},
  \bibinfo{author}{\bibfnamefont{C.}~\bibnamefont{Sun}},
  \bibinfo{author}{\bibfnamefont{D.}~\bibnamefont{Lohse}},
  \bibinfo{author}{\bibfnamefont{P.~A.} \bibnamefont{Tsai}}, \bibnamefont{and}
  \bibinfo{author}{\bibfnamefont{R.~G.~H.} \bibnamefont{Lammertink}},
  \bibinfo{journal}{PNAS} \textbf{\bibinfo{volume}{110}}, \bibinfo{pages}{8422}
  (\bibinfo{year}{2013}).

\bibitem[{\citenamefont{{Vinogradova} and
  {Yakubov}}(2003)}]{vinogradova.oi:2003}
\bibinfo{author}{\bibfnamefont{O.~I.} \bibnamefont{{Vinogradova}}}
  \bibnamefont{and} \bibinfo{author}{\bibfnamefont{G.~E.}
  \bibnamefont{{Yakubov}}}, \bibinfo{journal}{Langmuir}
  \textbf{\bibinfo{volume}{19}}, \bibinfo{pages}{1227} (\bibinfo{year}{2003}).

\bibitem[{\citenamefont{Cottin-Bizonne
  et~al.}(2005)\citenamefont{Cottin-Bizonne, Cross, Steinberger, and
  Charlaix}}]{charlaix.e:2005}
\bibinfo{author}{\bibfnamefont{C.}~\bibnamefont{Cottin-Bizonne}},
  \bibinfo{author}{\bibfnamefont{B.}~\bibnamefont{Cross}},
  \bibinfo{author}{\bibfnamefont{A.}~\bibnamefont{Steinberger}},
  \bibnamefont{and} \bibinfo{author}{\bibfnamefont{E.}~\bibnamefont{Charlaix}},
  \bibinfo{journal}{Phys. Rev. Lett.} \textbf{\bibinfo{volume}{94}},
  \bibinfo{pages}{056102} (\bibinfo{year}{2005}).

\bibitem[{\citenamefont{Vinogradova et~al.}(2009)\citenamefont{Vinogradova,
  Koynov, Best, and Feuillebois}}]{vinogradova.oi:2009}
\bibinfo{author}{\bibfnamefont{O.~I.} \bibnamefont{Vinogradova}},
  \bibinfo{author}{\bibfnamefont{K.}~\bibnamefont{Koynov}},
  \bibinfo{author}{\bibfnamefont{A.}~\bibnamefont{Best}}, \bibnamefont{and}
  \bibinfo{author}{\bibfnamefont{F.}~\bibnamefont{Feuillebois}},
  \bibinfo{journal}{Phys. Rev. Lett.} \textbf{\bibinfo{volume}{102}},
  \bibinfo{pages}{118302} (\bibinfo{year}{2009}).

\bibitem[{\citenamefont{Joly et~al.}(2006)\citenamefont{Joly, Ybert, and
  Bocquet}}]{joly.l:2006}
\bibinfo{author}{\bibfnamefont{L.}~\bibnamefont{Joly}},
  \bibinfo{author}{\bibfnamefont{C.}~\bibnamefont{Ybert}}, \bibnamefont{and}
  \bibinfo{author}{\bibfnamefont{L.}~\bibnamefont{Bocquet}},
  \bibinfo{journal}{Phys. Rev. Lett.} \textbf{\bibinfo{volume}{96}},
  \bibinfo{pages}{046101} (\bibinfo{year}{2006}).

\bibitem[{\citenamefont{Seo and Ducker}(2013)}]{seo.d:2013}
\bibinfo{author}{\bibfnamefont{D.}~\bibnamefont{Seo}} \bibnamefont{and}
  \bibinfo{author}{\bibfnamefont{W.~A.} \bibnamefont{Ducker}},
  \bibinfo{journal}{Phys. Rev. Lett.} \textbf{\bibinfo{volume}{111}},
  \bibinfo{pages}{174502} (\bibinfo{year}{2013}).

\bibitem[{\citenamefont{Quarteroni and Valli}(1999)}]{quarteroni.a:1999}
\bibinfo{author}{\bibfnamefont{A.}~\bibnamefont{Quarteroni}} \bibnamefont{and}
  \bibinfo{author}{\bibfnamefont{A.}~\bibnamefont{Valli}},
  \emph{\bibinfo{title}{Domain Decomposition Methods for Partial Differential
  Equations}} (\bibinfo{publisher}{Oxford Science Publications},
  \bibinfo{year}{1999}).

\bibitem[{\citenamefont{Pozrikidis}(1992)}]{Pozrikidis:book}
\bibinfo{author}{\bibfnamefont{C.}~\bibnamefont{Pozrikidis}},
  \emph{\bibinfo{title}{Boundary Integral and Singularity Methods for
  Linearised Viscous Flow}} (\bibinfo{publisher}{Cambridge University Press},
  \bibinfo{year}{1992}).

\bibitem[{\citenamefont{Zhou et~al.}(2013)\citenamefont{Zhou, Asmolov, Schmid,
  and Vinogradova}}]{Zhou_etal:2013}
\bibinfo{author}{\bibfnamefont{J.}~\bibnamefont{Zhou}},
  \bibinfo{author}{\bibfnamefont{E.~S.} \bibnamefont{Asmolov}},
  \bibinfo{author}{\bibfnamefont{F.}~\bibnamefont{Schmid}}, \bibnamefont{and}
  \bibinfo{author}{\bibfnamefont{O.~I.} \bibnamefont{Vinogradova}},
  \bibinfo{journal}{J. Chem. Phys.} \textbf{\bibinfo{volume}{139}},
  \bibinfo{pages}{174708} (\bibinfo{year}{2013}).

\bibitem[{\citenamefont{{Wang}}(2003)}]{wang2003}
\bibinfo{author}{\bibfnamefont{C.~Y.} \bibnamefont{{Wang}}},
  \bibinfo{journal}{Phys. Fluids} \textbf{\bibinfo{volume}{15}},
  \bibinfo{pages}{1114} (\bibinfo{year}{2003}).

\bibitem[{\citenamefont{Asmolov et~al.}(2013)\citenamefont{Asmolov, Zhou,
  Schmid, and Vinogradova}}]{asmolov_etal:2013}
\bibinfo{author}{\bibfnamefont{E.~S.} \bibnamefont{Asmolov}},
  \bibinfo{author}{\bibfnamefont{J.}~\bibnamefont{Zhou}},
  \bibinfo{author}{\bibfnamefont{F.}~\bibnamefont{Schmid}}, \bibnamefont{and}
  \bibinfo{author}{\bibfnamefont{O.~I.} \bibnamefont{Vinogradova}},
  \bibinfo{journal}{Phys. Rev. E} \textbf{\bibinfo{volume}{88}},
  \bibinfo{pages}{023004} (\bibinfo{year}{2013}).

\bibitem[{not()}]{note1}
\bibinfo{note}{We remark that this curve can be very well fitted by a symmetric
  fourth-order polynomial with slopes defnied by Eq.({\ref{slopes}}).}

\bibitem[{\citenamefont{de~Gennes}(1979)}]{degennes:1979}
\bibinfo{author}{\bibfnamefont{P.~G.} \bibnamefont{de~Gennes}},
  \bibinfo{journal}{C. R. Acad. Sci. Paris} \textbf{\bibinfo{volume}{288 B}},
  \bibinfo{pages}{219} (\bibinfo{year}{1979}).

\bibitem[{\citenamefont{Belyaev and Vinogradova}(2011)}]{belyaev.av:2011a}
\bibinfo{author}{\bibfnamefont{A.~V.} \bibnamefont{Belyaev}} \bibnamefont{and}
  \bibinfo{author}{\bibfnamefont{O.~I.} \bibnamefont{Vinogradova}},
  \bibinfo{journal}{Phys. Rev. Lett.} \textbf{\bibinfo{volume}{107}},
  \bibinfo{pages}{098301} (\bibinfo{year}{2011}).

\bibitem[{\citenamefont{Squires}(2008)}]{Squires08}
\bibinfo{author}{\bibfnamefont{T.~M.} \bibnamefont{Squires}},
  \bibinfo{journal}{Phys.~Fluids} \textbf{\bibinfo{volume}{20}},
  \bibinfo{pages}{092105} (\bibinfo{year}{2008}).

\bibitem[{\citenamefont{Bahga et~al.}(2010)\citenamefont{Bahga, Vinogradova,
  and Bazant}}]{bahga:2010}
\bibinfo{author}{\bibfnamefont{S.~S.} \bibnamefont{Bahga}},
  \bibinfo{author}{\bibfnamefont{O.~I.} \bibnamefont{Vinogradova}},
  \bibnamefont{and} \bibinfo{author}{\bibfnamefont{M.~Z.}
  \bibnamefont{Bazant}}, \bibinfo{journal}{J.~Fluid Mech.}
  \textbf{\bibinfo{volume}{644}}, \bibinfo{pages}{245} (\bibinfo{year}{2010}).

\bibitem[{\citenamefont{Huang et~al.}(2008)\citenamefont{Huang, Cottin-Bizzone,
  Ybert, and Bocquet}}]{Huang08}
\bibinfo{author}{\bibfnamefont{D.~M.} \bibnamefont{Huang}},
  \bibinfo{author}{\bibfnamefont{C.}~\bibnamefont{Cottin-Bizzone}},
  \bibinfo{author}{\bibfnamefont{C.}~\bibnamefont{Ybert}}, \bibnamefont{and}
  \bibinfo{author}{\bibfnamefont{L.}~\bibnamefont{Bocquet}},
  \bibinfo{journal}{Phys. Rev. Lett.} \textbf{\bibinfo{volume}{20}},
  \bibinfo{pages}{092105} (\bibinfo{year}{2008}).

\bibitem[{\citenamefont{Asmolov and Vinogradova}(2012)}]{asmolov:2012}
\bibinfo{author}{\bibfnamefont{E.~S.} \bibnamefont{Asmolov}} \bibnamefont{and}
  \bibinfo{author}{\bibfnamefont{O.~I.} \bibnamefont{Vinogradova}},
  \bibinfo{journal}{J. Fluid Mech.} \textbf{\bibinfo{volume}{706}},
  \bibinfo{pages}{108} (\bibinfo{year}{2012}).

\end{thebibliography}
}}

\end{document}